# Room-temperature high detectivity mid-infrared photodetectors based on black arsenic phosphorus

Mingsheng Long[1+], Anyuan Gao[1+], Peng Wang[2], Hui Xia[2], Claudia Ott[4], Chen Pan[1], Yajun Fu[1], Erfu Liu[1], Xiaoshuang Chen[2], Wei Lu[2], Tom Nilges[4], Jianbin Xu[5], Xiaomu Wang[3*], Weida Hu[2*], Feng Miao[1*]

**The mid-infrared (MIR) spectral range, pertaining to important applications such as molecular 'fingerprint' imaging, remote sensing, free space telecommunication and optical radar, is of particular scientific interest and technological importance. However, state-of-the-art materials for MIR detection are limited by intrinsic noise and inconvenient fabrication processes, resulting in high cost photodetectors requiring cryogenic operation. We report black arsenic-phosphorus-based long wavelength infrared photodetectors with room temperature operation up to 8.2 μm, entering the second MIR atmospheric transmission window. Combined with a van der Waals heterojunction, room temperature specific detectivity higher than $4.9 \times 10^9$ Jones was obtained in the 3-5 μm range. The photodetector works in a zero-bias photovoltaic mode, enabling fast photoresponse and low dark noise. Our van der Waals heterojunction photodector not only exemplify black arsenic-phosphorus as a promising candidate for MIR opto-electronic applications, but also pave the way for a general strategy to suppress 1/f noise in photonic devices.**

[1] National Laboratory of Solid State Microstructures, School of Physics, Collaborative Innovation Center of Advanced Microstructures, Nanjing University, Nanjing 210093, China.

[2] National Laboratory for Infrared Physics, Shanghai Institute of Technical Physics, Chinese Academy of Sciences, Shanghai 200083, China.

[3] School of Electronic Science and Technology, Nanjing University, Nanjing 210093, China.

[4] Synthesis and Characterization of innovative Materials, Technical University of Munich, Department of Chemistry, Garching b. München 85748, Germany

[5] Department of Electronic Engineering and Materials Science and Technology

Research Center, The Chinese University of Hong Kong, Hong Kong SAR

Correspondence and requests for materials should be addressed to F. M. (email: miao@nju.edu.cn), W. H. (email: wdhu@mail.sitp.ac.cn), or X. W. (email: xiaomu.wang@nju.edu.cn).

**Introduction**

State-of-the-art MIR detectors are generally made of certain narrow-band gap semiconductors, such as HgCdTe alloys(*1,2*) or quantum-well and quantum-dot structures based on group III–V materials(*3,4*). Unfortunately, these materials suffer from several major challenges that limit their wide application. First, the growth of such materials is usually sophisticated and environmentally hazardous, making it challenging to flexibly form heterojunction with other semiconductors. Second, operation of these detectors generally requires a cryogenic environment with complex cooling facilities, which prohibit their usage in portable contexts such as distributed environmental monitoring or compact telecommunication networks. The discovery of graphene has provided a promising alternative solution for MIR photodetectors(*5-8*) that can be easily fabricated and operated at room temperature. However, graphene's practical application has been limited by very low light absorption and an inherent vanishing band gap, which results in an extremely high dark current and noise level. High performance MIR photodetectors working at room temperature and in the second atmosphere window of long wavelength infrared (~8-14 μm) has yet to be demonstrated.

Very recently, a new two-dimensional (2D) layered material, black phosphorus, has arisen as an attractive candidate for opto-electronic applications(*9-15*), due to it is a tunable narrow band gap that is always direct regardless of layer number(*16-18*) and exhibits excellent strong in-plane anisotropic physical properties(*11,19-21*). In this work, we demonstrate high performance room-temperature MIR photodetectors based on black arsenic phosphorus (b-AsP), which is an alloy of black phosphorus with arsenic atoms in the forms of $As_{1-x}P_x$. By varying the composition of phosphorus, x, the band gap correspondingly changes from 0.3 eV to 0.15 eV. This energy range suggests that b-AsP may interact with light whose wavelength is as long as 8.5 μm. The extended detection range not only fully covers the first atmospheric window of mid-wavelength infrared (~3-5 μm) but is also broadened to the second atmospheric window of long-wavelength infrared (~8-14 μm), making b-AsP a highly attractive material for ultra-broadband photodetection and energy conversion.

**Results**

**Fabrication of b-AsP phototransistor.** We first examined the photo response of b-AsP by a phototransistor (as schematically shown in the inset of Fig. 1A). To prevent degradation of b-AsP flakes during the fabrication processes, we prepared b-AsP thin films by mechanically exfoliating bulk b-AsP samples (As$_{0.83}$P$_{0.17}$) onto a highly doped silicon substrate covered by 300 nm of SiO$_2$ in a glove box. We chose flakes of b-AsP ranging from 5 nm to 20 nm thick for device fabrication due to the desired compromise between high light absorption and low dark current. The devices are then fabricated by standard e-beam lithography, metallization and a lift-off process. After the fabrication processes, we spin-coated a thin layer of poly-methyl methacrylate (PMMA) to protect the samples from oxidation in the air. Figure 1A shows a typical optical absorption spectrum of the b-AsP samples we used. The absorption peak is located at approximately 2760 cm$^{-1}$, which corresponds to 3.62 μm. The relatively large thickness plays a beneficial role in boosting the optical absorption and thus the responsivity of the b-AsP-based photodetectors. As the wavenumber decreases from the peak, the absorption decreases linearly to approximately 1250 cm$^{-1}$ (corresponding to 8.27 μm), marked by the cross of two red lines (as guides to the eye). These results suggest that the absorption edge is at approximately 1250 cm$^{-1}$, corresponding to a ~0.15 eV band gap. Combined energy dispersive spectroscopy with Raman spectra studies, the compositions of the samples were confirmed to be As$_{0.83}$ P$_{0.17}$(*21*) (see Supplementary fig. S1). We also measured the electrical transport of b-AsP FETs, the field-effect mobility of which was calculated to be ~307 cm$^2$V$^{-1}$s$^{-1}$ at 0.01 V bias (see Supplementary fig. S2).

**Photocurrent mechanisms of b-AsP.** We now turn to study the origin of photocurrent generation in b-AsP. For simplicity and to clearly extract the intrinsic photoresponse, we still use the phototransistor structure. As shown in Fig. 1B, a typical phototransistor (shown in the down inset) exhibits nearly linear *I-V* curves under dark condition as well as under the illumination of an 8.05 μm MIR laser. The photocurrent ($I_P = I_{light} - I_{dark}$) increased linearly with increasing bias voltage, together with the observed zero-biased photoresponse (shown in of Fig. 1B upper inset) suggesting a significant MIR photoresponse. To reveal the detailed photoresponse mechanisms of the devices at the MIR range, we systematically measured the generated photocurrent at various source-drain voltage $V_{ds}$ and gate voltage $V_g$ values, with typical results presented in Fig. 1C and Fig. 1D. The photocurrent switched its polarity with increasing gate voltage at all source-drain biases. The opposite photocurrents in different regions are attributed to the photovoltaic effect (PVE) and photothermoelectric effect (PTE).

At low doped or intrinsic regimes (-15 V < $V_g$ < 15 V), the photocurrent is positive relative to $V_{ds}$, and reaches a maximum near the charge neutrality point. The positive polarity together with the zero-bias response suggests that the PVE dominates(*9,14,22-24*). In this scenario, the b-AsP/metal Schottky junction plays a key role in the photocurrent generation. Fig. 1E schematically illustrates the photovoltaic response of b-AsP devices in which photogenerated electron-hole pairs are separated at the b-AsP/metal junctions. If the channel is p-type doped, the photocurrent is mainly generated at the reverse-biased b-AsP/drain junction (top panel). In the case of slightly n-type doped b-AsP, the photocurrent is mainly generated at the reverse-biased b-AsP/source junction (bottom panel). Here, $V_{ds}$ is assumed to be positive regardless of the channel type for simplicity. In both cases, the photocurrent is positive relative to the conduction current. We further characterized these junctions through spatial photocurrent mapping measurements (at near infrared range). Supplementary fig. S3 presents the optical image of the device and corresponding photocurrent mapping results at $V_{ds}$ = 50 mV and 0 mV. The spatial mappings verify that the photocurrent is mainly generated at Schottky junctions. The photocurrent has opposite polarity at the two contacts due to the opposite junction bias direction. The mapping also excludes the photogating effect as the major working mechanism; if photogating dominates the response, the photocurrent would be mainly generated in the channel center. The PVE is mostly pronounced in the intrinsic regime due to the lower channel carrier density and longer photocarrier lifetime.

By contrast, the photocurrent is negative relative to $V_{ds}$ and shows very weak gate dependence in the highly doped regime ($V_g$ > 15 V), as shown in the line traces plotted in Fig. 1D. In this case, thermally driven processes (the PTE and bolometric effect) dominate the photoresponse. Unlike the case of graphene, for which the photocurrent is generated by the bolometric effect only, in b-AsP devices under low source-drain bias, the photocurrent is mainly attributed to the thermally driven processes due to the high electrical conductivity(*25-29*) and low thermal conductivity(*9,30*) of b-AsP. This result can be understood from the expression of the PTE-generated photocurrent: $I_{PTE} = (S_1 - S_2) \Delta T / R_d$, where $\Delta T$ is the temperature gradient, $R_d$ is the resistance of the device, and $S_1$ ($S_2$) is the Seebeck coefficient of b-AsP (metal electrodes). Under higher source-drain biases, the bolometric effect may be pronounced, manifested by a linearly increased photocurrent with $V_{ds}$. Compared with the PVE, the thermally driven processes present a lower responsivity. Therefore, we mainly operated our device in the PVE condition below.

**MIR photoresponse of b-AsP.** We next fully characterized the photoresponse of b-AsP in MIR. It is worth mentioning that the large photovoltaic response eventually generates photocurrent under zero source-drain bias through the unapparent asymmetry of metal electrodes or device shape. Higher photoreponses are expected in large built-in filed systems (e.g. device with asymmetric electrode or p-n junctions). Here for simplicity, we generally operated devices under zero bias, which also effectively suppresses dark current and therefore power consumption of photodetectors. We measured the zero-biased photoresponse of the b-AsP detectors at the MIR range from 2.4 μm to 8.05 μm, with typical responsivity data (defined as the ratio of photocurrent to incidence laser power) shown in Fig. 2A (see Supplementary fig. S4 for results from the visible (0.45 μm) to the near-infrared (1.55 μm) range). Although the responsivity decreases slightly with increasing wavelength of the illumination laser due to decreased optical absorption around the band edges, the device presented a high responsivity (15-30 mAW$^{-1}$) across the entire MIR range tested. To quantify the efficiency of light conversion to current, we extracted the external quantum efficiency (EQE), that is, the ratio of the number of photoexcited charge carriers to the number of incident photons. EQE can be derived by EQE = ($h c I_P / e \lambda P_I$), where $h$ is the Planck constant, $c$ is the speed of light, and $\lambda$ is the wavelength of the incident laser. The calculated EQE is as high as ~6.1% under the illumination of a 3.662 μm laser, which clearly indicates a promising performance at the MIR range. The speed of response is another important figure of merit of photodetectors, we further measured photoresponse time using a 4.034 μm infrared laser, with the results shown in Fig. 2B. The rise/fall time is defined as from 10/90% to 90/10% of the stable photocurrent after turning the laser on/off. The rise time, $\tau_{rise}$ = 0.54 ms, and the fall time, $\tau_{fall}$ = 0.52 ms, were obtained, as shown in Fig. 2B. Faster photoresponses at typical b-AsP FET devices were observed under the illumination of a 1.55 μm laser with higher power (see Supplementary fig. S5). In principle, a much faster photoresponse is expected for the PVE. We attribute the relatively slower photoresponse to the percolation transport mode resulting from the imperfect material interface. Namely, the electronic transport turns from hopping regime at low carrier densities to band-like regime at high carrier densities. As a result, the photoresponse is slower at low carrier densities (and thus for low incident light power) due to the low mobility and high disorder, which is consistent with Ref.12. Nevertheless, the speed demonstrated here is more than sufficient for IR imaging applications. We further measured the laser power dependent photocurrent near the absorption peak, with the calculated photoresponsivity and EQE plotted as a function of laser power, as shown in Fig. 2C. The measurements were performed under 3.662 μm MIR laser excitation at $V_{ds}$ = 0 V at room temperature. The photoresponsivity decreased from 180.0 mAW$^{-1}$ to 20.3 mAW$^{-1}$ as the power increased from 70.1 nW to 44.3 μW (the corresponding EQE decreased from 6.1% to 0.69%), which indicates that the photogating effect plays weaker roles in our illuminating power range due to the trapping centers being saturated under intense light(*12*).

The puckered crystal structure of b-AsP could naturally yield unique anisotropic photoresponses with many important applications, i.e., the photocurrent periodically varying with the polarization of incident light or the current collection direction. Fig. 2D shows the measured photocurrent along the x- (armchair edge) and y-direction (zigzag edge) of the same device (inset of Fig. 2D) at room temperature. The conductivity along the x-direction (without light illumination) is approximately 1.73 times higher than that along the y-direction at $V_g$ = 0 V. This anisotropic factor, $\sigma_{xx} / \sigma_{yy}$ = 1.73, is slightly larger than that in black phosphorus (~1.6)(*12*) and is consistent with previously reported results(*20,25,31*). Under the illumination of a 4.034 μm laser, $I_{Px}/I_{Py}$ is approximately 3.51 at $V_{ds}$ = 1 V. We also measured the polarization-resolved photoresponse, whereby the polarization of a linearly polarized incident laser was controlled by a half-wave plate. The polarization-dependent photocurrent mappings are presented in Supplementary fig. S6. The photocurrent was observed to be maximum when the light polarization was along the x-direction and minimum when the light was along the y-direction, similar to the observation in black phosphorus(*10,24*). The photocurrent anisotropy ratio, $γ = (I_{Pmax} - I_{Pmin}) / (I_{Pmax} + I_{Pmin})$, was estimated to be approximately 0.59, which is larger than that of black phosphorus (~0.3)(*24*).

**b-AsP based van der Waals photodetector.** High dark current noise is the major challenge of modern narrow-band gap semiconductor-based MIR photodetectors. Next, we demonstrate a general strategy to suppress dark current noise by using 2D van der Waals (vdW) heterojunctions. Integrability is an inherent merit of 2D materials, by which different 2D flakes can be sequentially stacked into vdW heterojunctions. Importantly, high energy barriers naturally formed at the interfaces of vdW junctions are able to effectively reduce dark noise. It should be noted that this highly desired yet facile strategy simply does not work in traditional materials owing to the difficulty of obtaining high-quality heterojunctions.

Following this idea, we fabricated photodetectors based on b-AsP/MoS$_2$ heterostructure. The photoresponse of a typical heterostructure device together with its optical image is shown in Fig. 3A. The b-AsP is a p-type semiconductor whereas MoS$_2$ is an n-type semiconductor. The typical rectification curves are presented in Fig. 3A, indicating that the van der Waals *p-n* junction was formed. This result is further confirmed by the photocurrent mapping at $V_{ds}$ = 0 V (see Supplementary fig. S7). The current at the forward bias is more than two orders of magnitude larger than that of under a reverse bias. Due to the energy barrier in the b-AsP/MoS$_2$ heterostructure, the dark current is markedly depressed. The photoresponsivity and EQE as a function of wavelength are plotted in fig. S8. The photoresponsivity ranges from 216.1 mAW$^{-1}$ to 115.4 mAW$^{-1}$ as the wavelength increased from 2.36 μm to 4.29 μm. The corresponding EQE decreased from 11.36% to 3.33%.

Finally, the current noise density was measured. As shown in Fig. 3B, the noise figure (and thus the detectivity, as discussed below) at the b-AsP/MoS$_2$ heterostructure was improved significantly compared with that of at the b-AsP FET. The frequency dependence results are also different, reflecting their different dominant noise sources. For the b-AsP FET devices, the 1/$f$ noise prevails at low frequencies (1-100 Hz) and is considerably above the Johnson noise level. 1/$f$ noise originates from fluctuations of local electronic states induced by the disorder or defects(*32*) which generally exist in 2D systems(*33,34*). Conversely, for the b-AsP/MoS$_2$ heterostructure devices, generation-recombination (g-r) noise dominates. G-r noise is caused by the fluctuation of carrier density due to the existence of trapping-detrapping centers. Its noise current spectrum is flat at low frequency and quickly decreases up to a frequency $f_0$. As shown in Fig. 3B, the spectrum fits a Lorentzian spectral model well(*35,36*), as:

$$\frac{<i_n^2>}{\Delta f} = \frac{A}{1+(f/f_0)^2}.$$

where $f_0 = 1/2\pi\tau$ is the 3 dB corner frequency and $\tau$ is the lifetime of the trap centers. These results clearly indicate that the energy barrier at the junction efficiently depresses the random transport of the photogenerated carriers and therefore inhibits the undesired 1/$f$ noise. Consequently, using the b-AsP/MoS$_2$ junction successfully decreased the total noise. Figure 3C presents the favorable noise equivalent power (NEP, defined by $i_n$ / $R$, $R$ as the responsivity and $i_n$ as the measured noise current) obtained in our devices. The room temperature NEP of a junction at the MIR range is below 0.24 pWHz$^{-1/2}$, and that of FET is lower than 4.35 pWHz$^{-1/2}$, even for 8.05 μm MIR light. With the knowledge of noise density and NEP, another important figure of merit is the specific detectivity, $D^*$, which determines the minimum illumination light power that a detector can distinguish from the noise. This value can be calculated by $D^* = (A\,B)^{1/2}$ / NEP where $A$ is the active area of the device and $B$ is the measuring bandwidth. The active area is used to normalize the dark noise. Figure 3D presents $D^*$ as a function of wavelength. For comparison, data from the best available room temperature operated MIR semiconductor (PbSe-based) detector, bolometer and thermopile are also given in the figure(*37,38*). Remarkably, the peak $D^*$ of our junction approaches 9.2×10$^9$ Jones and it is consistently larger than $4.9\times10^9$ Jones in the 3-5 μm range; these values are well beyond all room temperature MIR photodetectors to date (e.g., the black line in Fig. 3D). Actually, the room temperature $D^*$ of b-AsP FET is considerably larger than $1.06\times10^8$ Jones ($cmHz^{1/2}W^{-1}$), even for 8.05 μm MIR light, which is already higher than that of the commercial thermistor bolometer (the purple line in Fig. 3D). The performance of photodetection shows a significant enhancement for these b-AsP/MoS$_2$ heterostructure devices.

**Discussion**

In summary, we demonstrated room temperature operated MIR (entering the second atmospheric transmission window) photodetectors based on b-AsP. Compared with other MIR detectors, such as graphene(*5,39,40*), the b-AsP detectors exhibit significant advantages of promising broadband MIR responsivity, fast speed and excellent anisotropic photoresponse. In addition, the 2D nature of b-AsP renders it inherently easy to integrate with other materials. The specific detectivity is one of the most important figure of merit for photodetectors. Long wavelength detection generally requires small gap semiconductors to absorb light. For junctionless photoconductors, especially for the narrow band gap 2D materials, poor dark noise causes low signal to noise ratio and small specific detectivity. Junctions thus propose an effective approach to enhance specific detectivity considerably. Taking benefit from the promising optical properties of b-AsP and facile fabrication of van der Waals heterojunctions, we demonstrated that the overall performances, especially the dark current noise and specific detectivity can be further improved. The main working mechanisms of the devices were also revealed. Further work may include large area synthesis of b-AsP thin films and scalable fabrication of MIR devices. Our findings not only exemplify an ideal photodetector for challenging MIR imaging tasks but also pave the way for novel MIR technologies, such as polarization sensitive detection and free space telecommunication.

**Materials and Methods**
**Materials synthesis**

Bulk black arsenic phosphorus (b-$As_xP_{1-x}$) crystals were synthesized using the mineralizer-assisted short-way transport reaction method(*41*). Briefly, a mixture of gray arsenic and red phosphorus with molar ratios ranging from 5:5 to 2:8 was used as the precursor. Pre-synthesized lead iodide ($PbI_2$, weighing 10 mg per 500 mg) was added as the mineralization agent. The mixture was then evacuated in a 10 cm silica glass ampoule and placed horizontally in a furnace. The mixture was heated up to 550 °C for 8 h, held at this temperature for 20 to 80 h and slowly cooled to room temperate within 20 h. In this process, the heating elements of the furnace were configured within the walls. The mixture of reactive materials was located at the hot end, with the empty part of the ampoule towards the cooler center. The arsenic composition, x, obtained by this method is distributed from 0.36 to 0.83. The b-AsP samples with different arsenic compositions were tested in this project, and a typical set of results from the sample with x~0.83 is described in the main text.

**Device fabrication**

We used a standard mechanical exfoliation method to isolate few-layer black phosphorus flakes, typically 5-20 nm, on a highly doped Si wafer covered by a 300-nm-thick $SiO_2$ layer. The thickness of the flakes was first measured using a Bruker Multimode 8 atomic force microscope (AFM). The b-AsP/$MoS_2$ heterostructure was fabricated using a polymer-free van der Waals assembly technique in a glove box filled with an inert atmosphere. The devices were fabricated using a conventional electron-beam lithography process followed by standard electron-beam evaporation of metal electrodes (typically 5 nm Ti/ 50 nm Au).

After the fabrication processes, we spin-coated a thin layer of poly-methyl methacrylate (PMMA) to protect the samples from oxidation in air. The stability improvement was verified by checking the optical image, dark current and photovoltaic response (see Supplementary fig. S9). We did not find any obvious degradation in those protected samples fabricated two months ago.

**Electrical and photoresponse measurements**

Electrical transport and photoresponse measurements were performed using a Keithley 2636A dual channel digital source meter. The wavelength-dependent photoresponse in Fig. 2A was measured using a custom-built wavelength tunable multichannel MIR laser source. The spectrum spans from 2 μm to 4.3 μm with ~0.43 mm$^2$ spot size. The 5.3 μm and 8.05 μm light sources are custom-built quantum cascade lasers with ~9 mm$^2$ spot size and ~50 mW power. In the visible to near infrared range from 450 nm to 1550 nm, the laser was focused on the device using a 20× objective lens. Noise measurements were performed at room temperature. The devices were set in a thoroughly screened metal box to ensure that the device was working in the dark and to reduce the noise originating from the environment. Noise spectra were acquired by a spectrum analyzer (Stanford Research System SR770, with 100kHz measuring bandwidth) at different biases. All the measurements were performed under ambient conditions.

**Supplementary Materials**

- fig. S1. Raman spectra of b-AsP with different thicknesses.
- fig. S2. The transfer curves of two typical b-AsP FET devices.
- fig. S3. The photocurrent mappings of a typical device at near infrared range.
- fig. S4. The performance of a typical b-AsP device at visible and near infrared range.
- fig. S5. Fast photoresponse at near infrared.
- fig. S6. Laser polarization direction sensitive photocurrent mapping.
- fig. S7. Photocurrent mapping of the b-As$_{0.83}$P$_{0.17}$/MoS$_2$ heterostructure.
- fig. S8. Photoresponsivity and EQE of a typical b-As$_{0.83}$P$_{0.17}$/MoS$_2$ heterostructure device.
- fig. S9. The stability of b-AsP samples spin coated by a PMMA layer.

**References and Notes**

# References:

1. P. Norton. HgCdTe infrared detectors. *Opto-Electron. Rev.* **10**, 159-174 (2002).

2. A. Rogalski. Toward third generation HgCdTe infrared detectors. *J. Alloy. Compd.* **371**, 53-57 (2004).
3. B.F. Levine, K.K. Choi, C.G. Bethea, J. Walker & R.J. Malik. New 10 μm infrared detector using intersubband absorption in resonant tunneling GaAlAs superlattices. *Appl. Phys. Lett.* **50**, 1092 (1987).
4. A. Rogalski & P. Martyniuk. InAs/GaInSb superlattices as a promising material system for third generation infrared detectors. *Infrared Phys. Techn.* **48**, 39-52 (2006).
5. B.Y. Zhang, T. Liu, B. Meng, X. Li, G. Liang, X. Hu & Q.J. Wang. Broadband high photoresponse from pure monolayer graphene photodetector. *Nat. Commun.* **4**, 1811 (2013).
6. M. Long, E. Liu, P. Wang, A. Gao, H. Xia, W. Luo, B. Wang, J. Zeng, Y. Fu, K. Xu, W. Zhou, Y. Lv, S. Yao, M. Lu, Y. Chen, Z. Ni, Y. You, X. Zhang, S. Qin, Y. Shi, W. Hu, D. Xing & F. Miao. Broadband Photovoltaic Detectors Based on an Atomically Thin Heterostructure. *Nano Lett.* **16**, 2254-2259 (2016).
7. G. Konstantatos, M. Badioli, L. Gaudreau, J. Osmond, M. Bernechea, F.P. Garcia De Arquer, F. Gatti & F.H.L. Koppens. Hybrid graphene-quantum dot phototransistors with ultrahigh gain. *Nature Nanotech.* **7**, 363-368 (2012).
8. L. Chang-Hua, C. You-Chia, T.B. Norris & Z. Zhaohui. Graphene photodetectors with ultra-broadband and high responsivity at room temperature. *Nature Nanotech.* **9**, 273-278 (2014).
9. N. Youngblood, C. Chen, S.J. Koester & M. Li. Waveguide-integrated black phosphorus photodetector with high responsivity and low dark current. *Nature Photon.* **9**, 247-252 (2015).
10. H. Yuan, X. Liu, F. Afshinmanesh, W. Li, G. Xu, J. Sun, B. Lian, A.G. Curto, G. Ye, Y. Hikita, Z. Shen, S. Zhang, X. Chen, M. Brongersma, H.Y. Hwang & Y. Cui. Polarization-sensitive broadband photodetector using a black phosphorus vertical p-n junction. *Nature Nanotech.* **10**, 707-713 (2015).
11. J. Qiao, X. Kong, Z. Hu, F. Yang & W. Ji. High-mobility transport anisotropy and linear dichroism in few-layer black phosphorus. *Nat. Commun.* **5**, 4475 (2014).
12. Q. Guo, A. Pospischil, M. Bhuiyan, H. Jiang, H. Tian, D. Farmer, B. Deng, C. Li, S. Han, H. Wang, Q. Xia, T. Ma, T. Mueller & F. Xia. Black Phosphorus Mid-Infrared Photodetectors with High Gain. *Nano Lett.* **16**, 4648-4655 (2016).
13. S. Zhang, J. Yang, R. Xu, F. Wang, W. Li, M. Ghufran, Y. Zhang, Z. Yu, G. Zhang, Q. Qin & Y. Lu. Extraordinary Photoluminescence and Strong Temperature/Angle-Dependent Raman Responses in Few-Layer Phosphorene. *ACS Nano* **8**, 9590-9596 (2014).
14. T. Low, A.S. Rodin, A. Carvalho, Y. Jiang, H. Wang, F. Xia & A.H.C. Neto. Tunable optical properties of multilayer black phosphorus thin films. *Phys. Rev. B* **90**, 75434 (2014).
15. M. Buscema, D.J. Groenendijk, G.A. Steele, H.S.J. van der Zant & A. Castellanos-Gomez. Photovoltaic effect in few-layer black phosphorus PN junctions defined by local electrostatic gating. *Nat. Commun.* **5**, 4651 (2014).
16. A.S. Rodin, A. Carvalho & A.H. Castro Neto. Strain-Induced Gap Modification in Black Phosphorus. *Phys. Rev. Lett.* **112**, 176801 (2014).
17. G. Jie, Z. Zhen & D. Tomanek. Phase Coexistence and Metal-Insulator Transition in Few-Layer Phosphorene: A Computational Study. *Phys. Rev. Lett.* **113**, 46804 (2014).
18. S. Das, W. Zhang, M. Demarteau, A. Hoffmann, M. Dubey & A. Roelofs. Tunable Transport Gap in Phosphorene. *Nano Lett.* **14**, 5733-5739 (2014).
19. X. Wang, A.M. Jones, K.L. Seyler, V. Tran, Y. Jia, H. Zhao, H. Wang, L. Yang, X. Xu & F. Xia. Highly anisotropic and robust excitons in monolayer black phosphorus. *Nature Nanotech.* **10**, 517-521 (2015).
20. F. Xia, H. Wang & Y. Jia. Rediscovering black phosphorus as an anisotropic layered material for

optoelectronics and electronics. *Nat. Commun.* **5**, 4458 (2014).

21. B. Liu, M. Köpf, A.N. Abbas, X. Wang, Q. Guo, Y. Jia, F. Xia, R. Weihrich, F. Bachhuber, F. Pielnhofer, H. Wang, R. Dhall, S.B. Cronin, M. Ge, X. Fang, T. Nilges & C. Zhou. Black Arsenic-Phosphorus: Layered Anisotropic Infrared Semiconductors with Highly Tunable Compositions and Properties. *Adv. Mater.* **27**, 4423-4429 (2015).

22. M. Freitag, T. Low, F. Xia & P. Avouris. Photoconductivity of biased graphene. *Nature Photon.* **7**, 53-59 (2012).

23. M. Buscema, J.O. Island, D.J. Groenendijk, S.I. Blanter, G.A. Steele, H.S.J. van der Zant & A. Castellanos-Gomez. Photocurrent generation with two-dimensional van der Waals semiconductors. *Chem. Soc. Rev.* **44**, 3691-3718 (2015).

24. T. Hong, B. Chamlagain, W. Lin, H. Chuang, M. Pan, Z. Zhou & Y. Xu. Polarized photocurrent response in black phosphorus field-effect transistors. *Nanoscale* **6**, 8978-8983 (2014).

25. H. Liu, A.T. Neal, Z. Zhu, Z. Luo, X. Xu, D. Tománek & P.D. Ye. Phosphorene: An Unexplored 2D Semiconductor with a High Hole Mobility. *ACS Nano* **8**, 4033-4041 (2014).

26. L. Li, Y. Yu, G.J. Ye, Q. Ge, X. Ou, H. Wu, D. Feng, X.H. Chen & Y. Zhang. Black phosphorus field-effect transistors. *Nature Nanotech.* **9**, 372-377 (2014).

27. L. Li, G.J. Ye, V. Tran, R. Fei, G. Chen, H. Wang, J. Wang, K. Watanabe, T. Taniguchi, L. Yang, X.H. Chen & Y. Zhang. Quantum oscillations in a two-dimensional electron gas in black phosphorus thin films. *Nature Nanotech.* **10**, 608-613 (2015).

28. X. Chen, Y. Wu, Z. Wu, Y. Han, S. Xu, L. Wang, W. Ye, T. Han, Y. He, Y. Cai & N. Wang. High-quality sandwiched black phosphorus heterostructure and its quantum oscillations. *Nat. Commun.* **6**, 7315 (2015).

29. N. Gillgren, D. Wickramaratne, Y. Shi, T. Espiritu, J. Yang, J. Hu, J. Wei, X. Liu, Z. Mao, K. Watanabe, T. Taniguchi, M. Bockrath, Y. Barlas, R.K. Lake & C.N. Lau. Gate tunable quantum oscillations in air-stable and high mobility few-layer phosphorene heterostructures. *2D Mater.* **2**, 110011 (2015).

30. Z. Luo, J. Maassen, Y. Deng, Y. Du, R.P. Garrelts, M.S. Lundstrom, P.D. Ye & X. Xu. Anisotropic in-plane thermal conductivity observed in few-layer black phosphorus. *Nat. Commun.* **6**, 8572 (2015).

31. A. Mishchenko, Y. Cao, G.L. Yu, C.R. Woods, R.V. Gorbachev, K.S. Novoselov, A.K. Geim & L.S. Levitov. Nonlocal Response and Anamorphosis: The Case of Few-Layer Black Phosphorus. *Nano Lett.* **15**, 6991-6995 (2015).

32. N. Clément, K. Nishiguchi, A. Fujiwara & D. Vuillaume. One-by-one trap activation in silicon nanowire transistors. *Nat. Commun.* **1**, 92 (2010).

33. J. Na, Y.T. Lee, J.A. Lim, D.K. Hwang, G. Kim, W.K. Choi & Y. Song. Few-Layer Black Phosphorus Field-Effect Transistors with Reduced Current Fluctuation. *ACS Nano* **8**, 11753-11762 (2014).

34. A.A. Balandin. Low-frequency 1/f noise in graphene devices. *Nature Nanotech.* **8**, 549-555 (2013).

35. L.D.Y.A. SAH. Theory and Experiments of Low-Frequency Generation-Recombination Noise in MOS Transistors. *IEEE T. Electron Dev.* **ED-16**, 170-177 (1969).

36. J.A. COPELAND. Semiconductor Impurity Analysis from Low-Frequency Noise Spectra. *IEEE T. Electron Dev.* **ED-18**, 50-53 (1971).

37. http://www.vision-systems.com/articles/print/volume-16/issue-4/features/the-infrared-choice.html.

38. A.L. Hsu, P.K. Herring, N.M. Gabor, S. Ha, Y.C. Shin, Y. Song, M. Chin, M. Dubey, A.P. Chandrakasan, J. Kong, P. Jarillo-Herrero & T. Palacios. Graphene-Based Thermopile for Thermal

Imaging Applications. *Nano Lett.* **15**, 7211-7216 (2015).

39. X. Wang, Z. Cheng, K. Xu, H.K. Tsang & J. Xu. High-responsivity graphene/silicon-heterostructure waveguide photodetectors. *Nature Photon.* **7**, 888-891 (2013).

40. Y. Yao, R. Shankar, P. Rauter, Y. Song, J. Kong, M. Loncar & F. Capasso. High-Responsivity Mid-Infrared Graphene Detectors with Antenna-Enhanced Photocarrier Generation and Collection. *Nano Lett.* **14**, 3749-3754 (2014).

41. O. Osters, T. Nilges, F. Bachhuber, F. Pielnhofer, R. Weihrich, M. Schöneich & P. Schmidt. Synthesis and Identification of Metastable Compounds: Black Arsenic-Science or Fiction? *Angew. Chem. Int. Ed.* **51**, 2994-2997 (2012).

**Acknowledgments**

**Funding:**

This work was supported in part by the National Key Basic Research Program of China (2015CB921600, 2013CBA01603, 2013CB632700), the National Natural Science Foundation of China (61625402, 11374142, 61674157, 61574076), the Natural Science Foundation of Jiangsu Province (BK20140017, BK20150055), the Fund of the Shanghai Science and Technology Foundation (14JC1406400), the Specialized Research Fund for the Doctoral Program of Higher Education (20130091120040), and Fundamental Research Funds for the Central Universities and the Collaborative Innovation Center of Advanced Microstructures.

**Author contribution:**

X. W., F. M., W. H and M. L. conceived the project and designed the experiments. M. L., A. G., P. W., and H. X. performed device fabrication and characterization. M. L., A. G., X. W., F. M. and W. H. performed data analysis and interpretation. X. W., M. L., F. M. and W. H co-wrote the paper, and all authors contributed to the discussion and preparation of the manuscript. M. L. and A. G. contributed equally to this work.

**Competing financial interests:**

The authors declare no competing financial interests.

**Data and materials availability:** All data needed to evaluate the conclusions in the paper are present in the paper and/or the Supplementary Materials. Additional data available from authors upon request.

Fig. 1. MIR photovoltaic detector based on b-AsP. (A) Infrared absorption spectra of the b-As$_{0.83}$P$_{0.17}$ sample. The inset is the schematic drawing of the b-As$_{0.83}$P$_{0.17}$ phototransistor for photodetection. (B), $I_{ds}$-$V_{ds}$ characteristic curves with and without illumination, and photocurrent $I_P$ as a function of bias voltage at $V_g$ = 0 V. The wavelength of the laser was 8.05 μm and the power density was 0.17 Wcm$^{-2}$. Inset: optical image of the device. The scale bar is 5 μm. (C) Two-dimensional counter plot of the MIR (4.034 μm) photocurrent as a function of $V_{ds}$ and $V_g$. The photocurrent generation mechanism is dominated by the PVE and PTE at zero bias voltage. The incident laser power density was fixed at ~ 0.1 Wcm$^{-2}$. (D) Photocurrent versus gate voltage at various bias voltages. The sign of the photocurrent changes as the gate voltage increases at ~15 V from negative (p-doped) to positive (highly n-doped). (E) Schematic diagrams of energy structure diagrams at different doping types under a bias voltage $V_{ds}$. Top panel: the sample of b-AsP working at the p-type region. Bottom panel: the device working at the n-type region. The black horizontal arrows indicate the direction of the photocurrent, which was caused by the PVE.

Fig. 2. Performance of the b-AsP photodetectors at MIR range at room temperature. (A) Photoresponsivity $R$ (left) and EQE (right) of a typical device for wavelengths ranging from 2.4 μm to 8.05 μm. The measurements were carried out at $V_{ds}$ = 0 V and $V_g$ = 0 V. (B) Fast photoresponse of a typical device measured under a 4.034 μm laser (21.5 Wcm$^{-2}$) at $V_{ds}$ = 0 V and $V_g$ = 0 V. Here, the rise/fall time was defined as the photocurrent increased/decreased from 10/90% to 90/10% of the stable photocurrent. (C) Measured photoresponsivity $R$ (left axis) and external quantum efficiency EQE (right axis) of a typical device versus power of the incident laser (4.034 μm). The measurements were performed with $V_{ds}$ = 0 V and $V_g$ = 0 V. (D) The $I_{ds}$-$V_{ds}$ curves with and without illumination of the device. The x- and y- directions are labeled in the optical image in the inset. The scale bar is 5 μm. The wavelength of the incident

laser was 4.034 μm and the laser power was fixed at 21.5 Wcm$^{-2}$.

**Fig. 3. Rectifying curves and photoresponse of the b-AsP/MoS$_2$ heterostructure detectors.** (**A**) $I_{ds}$-$V_{ds}$ characteristic curves (in logarithmic scale) with and without illumination ($V_g$ = 0 V). The wavelength of the laser was 4.034 μm and the power density was 1.09 Wcm$^{-2}$. The inset: the optical image of a typical b-AsP/MoS$_2$ heterostructure device. The scale bar is 5 μm. (**B**) The current noise power spectra at $V_{ds}$ = 0 V of a b-AsP FET device (blue open circles) and a b-AsP/MoS$_2$ heterostructure (red open squares). The black solid line (plotted as $A/[1+(f/f_0)^2]$) is a reference for the 1/$f$ noise trend. (**C**) Wavelength dependence of the noise equivalent power. (D) Wavelength dependence of the specific detectivity, $D^*$ (right axis), at $V_{ds}$ = 0 V. The purple and dark lines are commercial specific detectivity for a thermistor bolometer and PbSe MIR detectors, respectively, at room temperature. The error bar of b-AsP FET stands for the non-uniformity of photocurrent across the channel.

# Figure 1

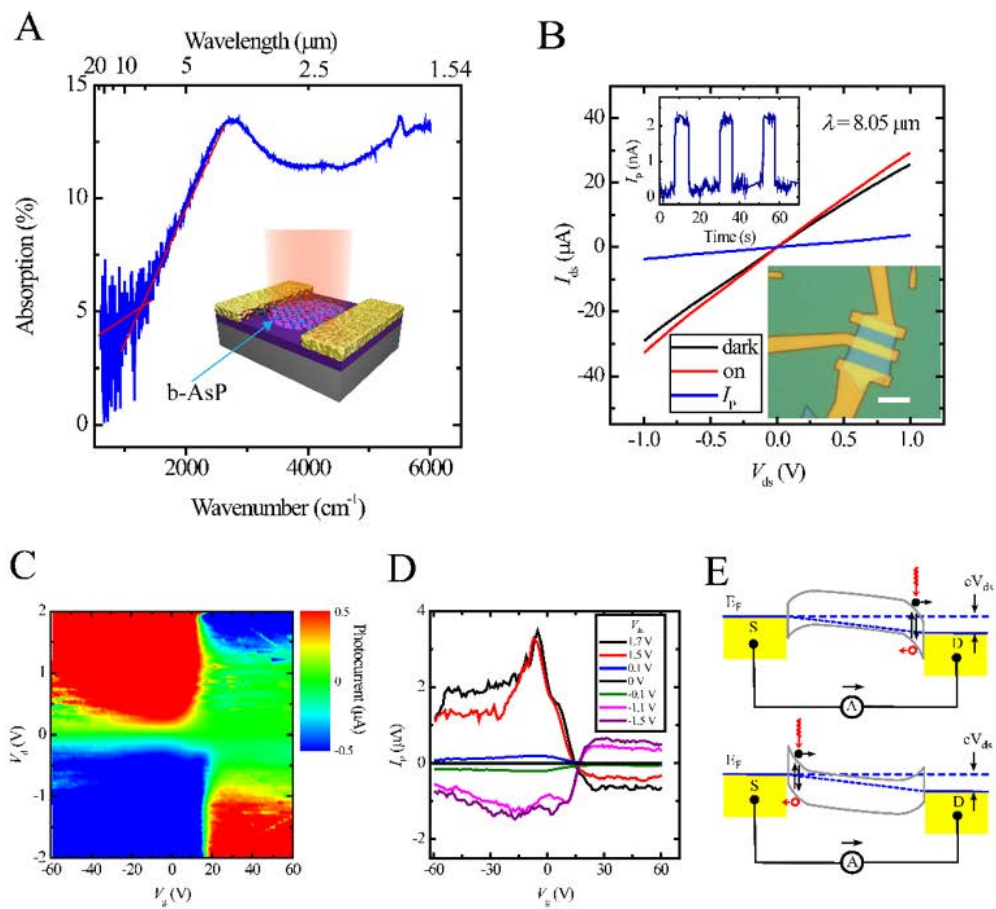

# Figure 2

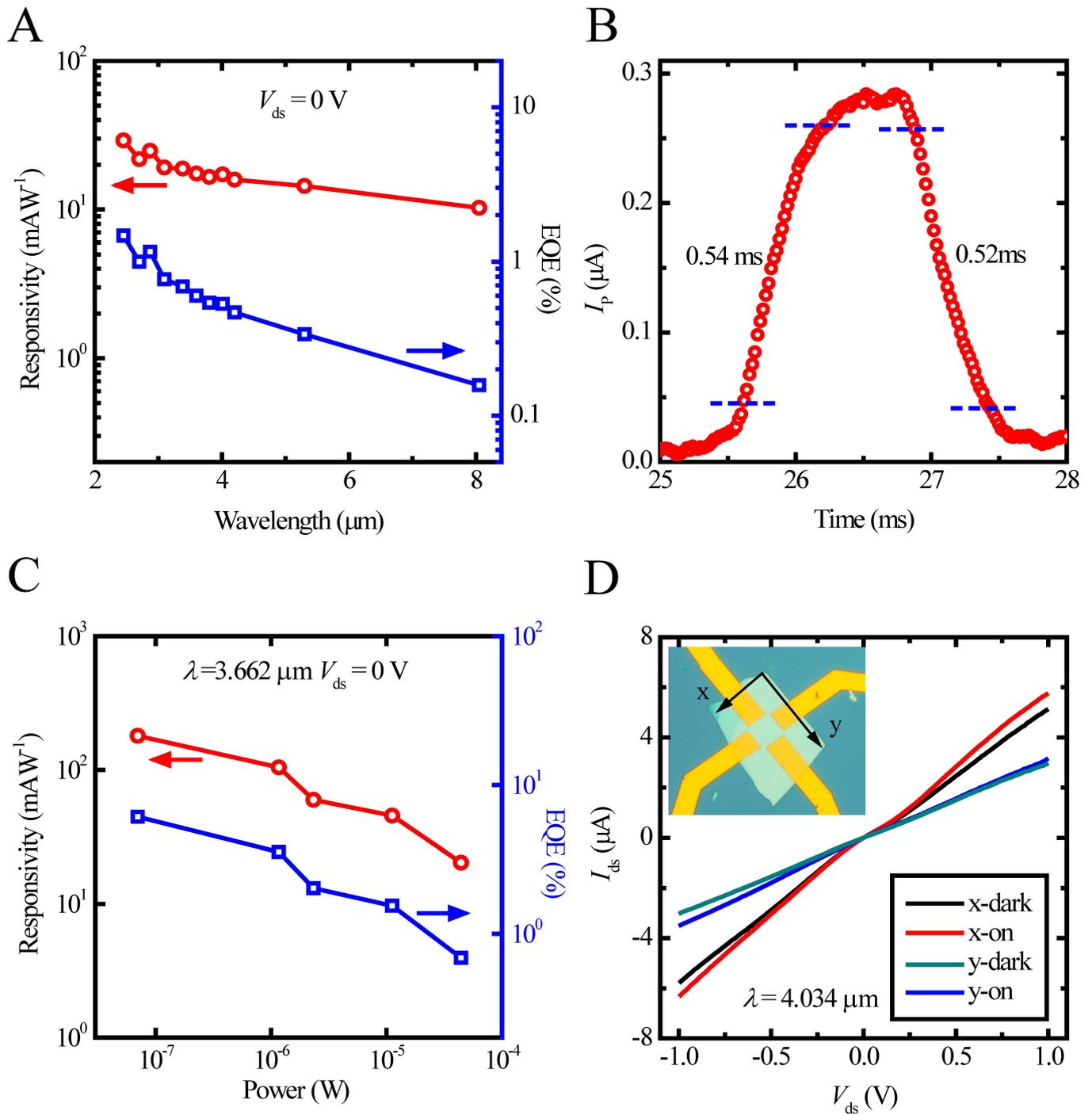

# Figure 3

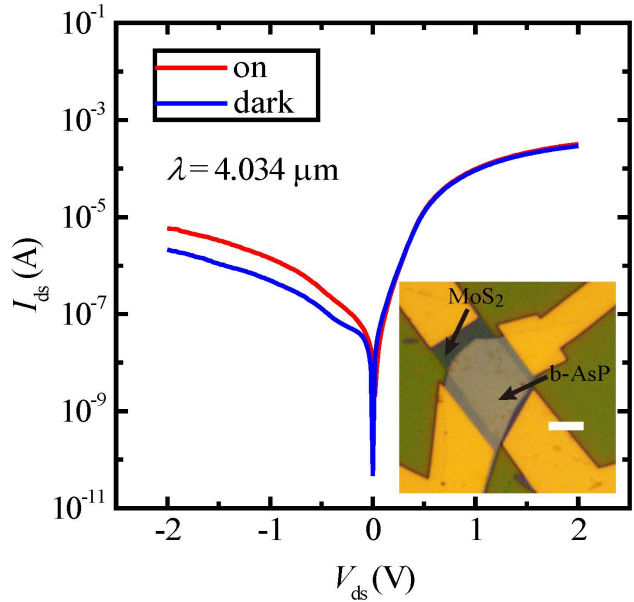
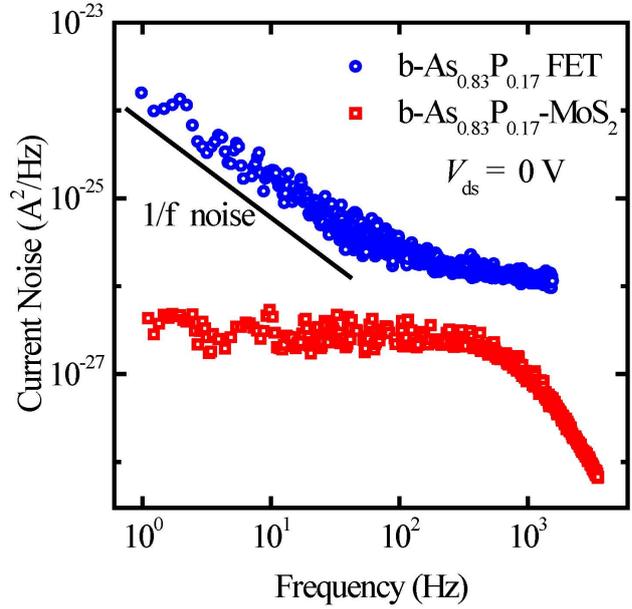
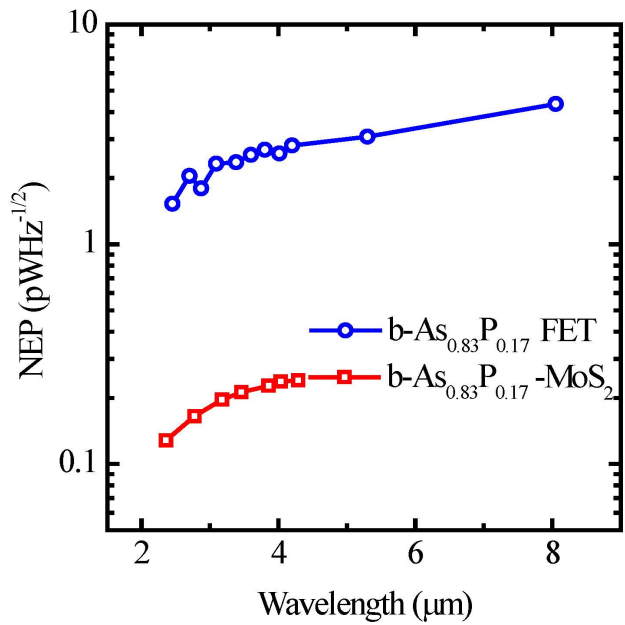
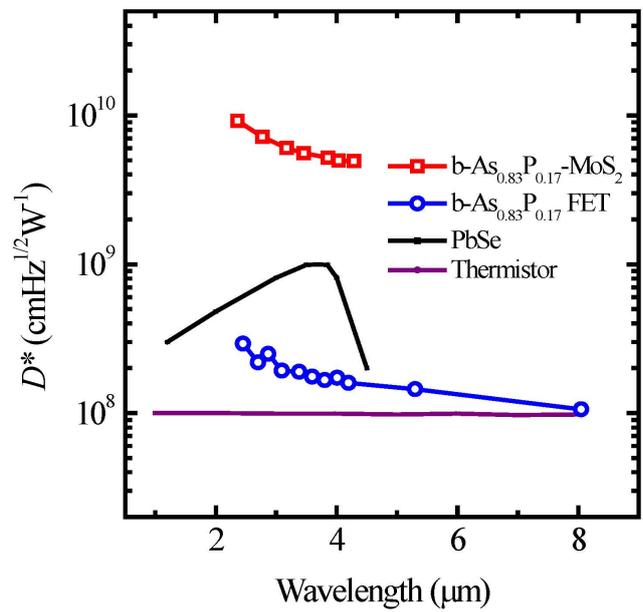

Supplementary Materials for

**Room-temperature high detectivity mid-infrared photodetectors based on black arsenic phosphorus**

Mingsheng Long[1+], Anyuan Gao[1+], Peng Wang[2], Hui Xia[2], Claudia Ott[4], Chen Pan[1], Yajun Fu[1], Erfu Liu[1], Xiaoshuang Chen[2], Wei Lu[2], Tom Nilges[4], Jianbin Xu[5], Xiaomu Wang[3*], Weida Hu[2*], Feng Miao[1*]

## This PDF file includes:

- fig. S1. Raman spectra of b-AsP with different thicknesses.
- fig. S2. The transfer curves of two typical b-AsP FET devices.
- fig. S3. The photocurrent mappings of a typical device at near infrared range.
- fig. S4. The performance of a typical b-AsP device at visible and near infrared range.
- fig. S5. Fast photoresponse at near infrared.
- fig. S6. Laser polarization direction sensitive photocurrent mapping.
- fig. S7. Photocurrent mapping of the b-As$_{0.83}$P$_{0.17}$/MoS$_2$ heterostructure.
- fig. S8. Photoresponsivity and EQE of a typical b-As$_{0.83}$P$_{0.17}$/MoS$_2$ heterostructure device.
- fig. S9. The stability of b-AsP samples spin coated by a PMMA layer.

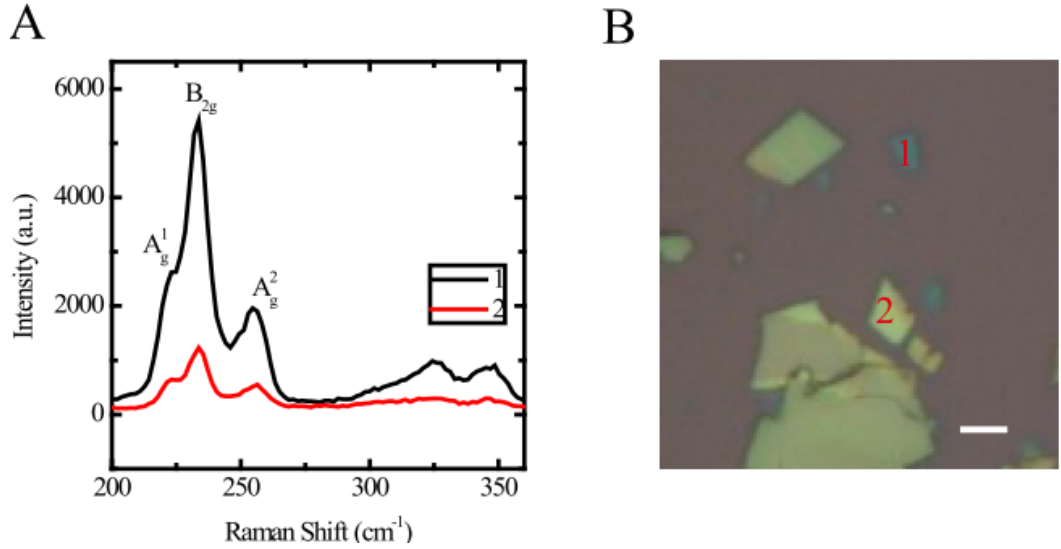

**fig. S1. Raman spectra of b-AsP with different thicknesses.** (**A**) Three characteristic vibrational modes: $A_g^1$, $B_{2g}$ and $A_g^2$, located at 224 cm$^{-1}$, 233cm$^{-1}$ and 256 cm$^{-1}$, respectively. Numbers 1 and 2 correspond to different flakes with optical images shown in (B). The silicon peak at 520 cm$^{-1}$ was used as calibration of the measurements. (**B**) Optical image of the sample for Raman measurement with the scale bar of 5 μm. A 50× objective lens was used to focus the 514 nm laser, with beam size ~1 μm.

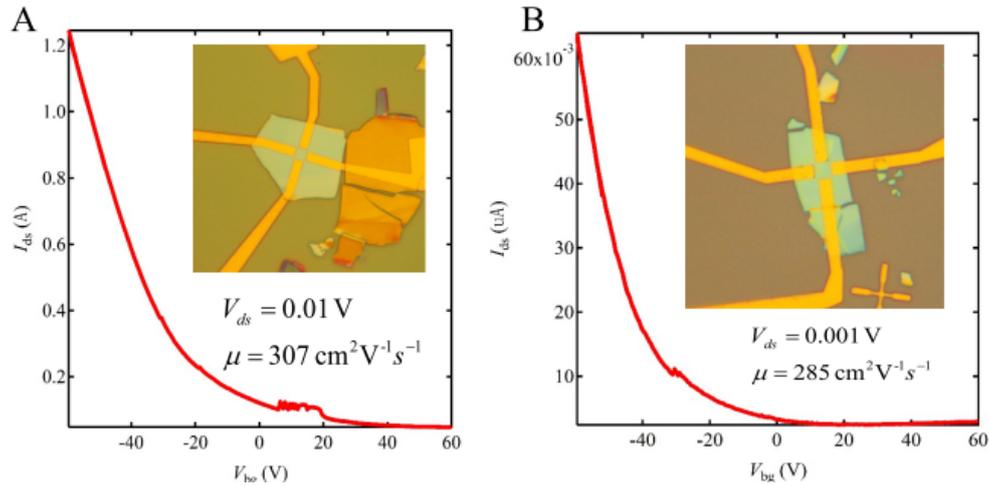

**fig. S2. The transfer curves of two typical b-AsP FET devices.** The length-to-width ratio of the channels is ~1. The mobility of device#1 (A) and device #2 (B) was estimated to be ~307 cm$^2$V$^{-1}$s$^{-1}$ and ~285 cm$^2$V$^{-1}$s$^{-1}$, respectively.

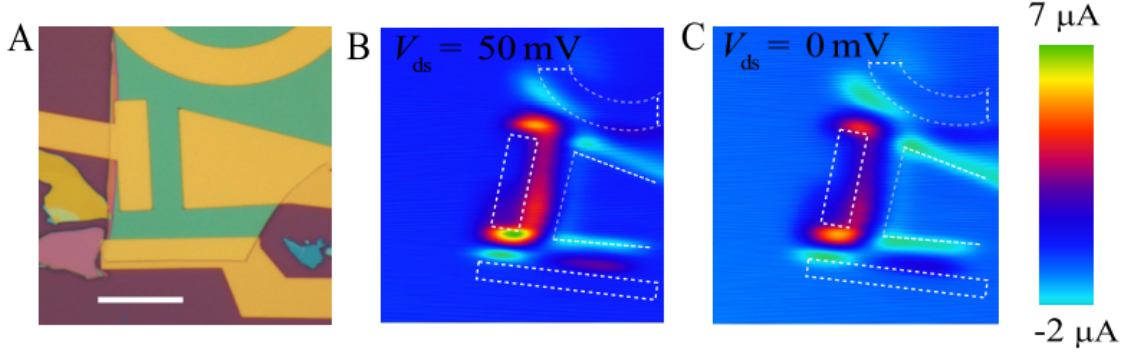

**fig. S3. The photocurrent mappings of a typical device at near infrared range.** (**A**) The optical image of a b-As$_{0.83}$P$_{0.17}$ FET device for photocurrent mapping study. The scale bar is 5 μm. (**B**) and (**C**) The photocurrent mappings measured with source-drain bias of 50 mV and 0 mV respectively. The wavelength of the laser was 1060 nm and the power was fixed at 64.5 μW.

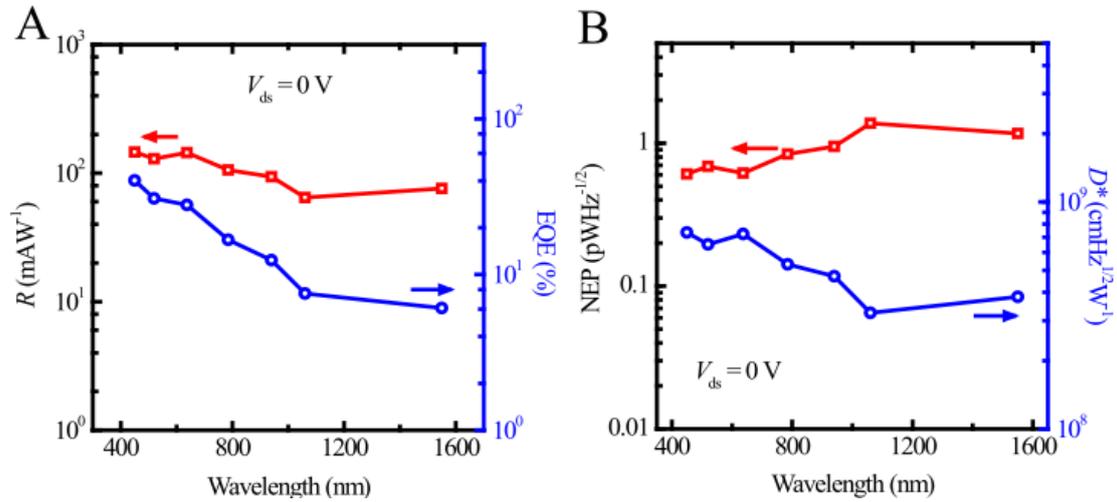

**fig. S4. The performance of a typical b-AsP device at visible and near infrared range.** (**A**) Photoresponsivity $R$ (left axis) and EQE (right) of a typical device for wavelength ranging from 450 nm to 1550 nm. The photoresponsivity and EQE were measured to be up to 146.0 mAW$^{-1}$ and 40.2% respectively under a 450 nm laser illumination at $V_g$ = 0 V. (**B**) Wavelength dependence of the noise equivalent power (NEP, left axis) and specific detectivity $D^*$ (right) at $V_{ds}$ = 0 V for wavelength ranging from 450 nm to 1550 nm. The NEP is lower than 1.3 pWHz$^{-1/2}$, indicating the detector is capable of detecting weak light signal below 1.3 pW at visible and near infrared range. The specific detectivity $D^*$ is up to 7.33 ×10$^8$ Jones for 450 nm and 3.8×10$^8$ Jones for 1550 nm laser. The measurements were performed at ambient conditions.

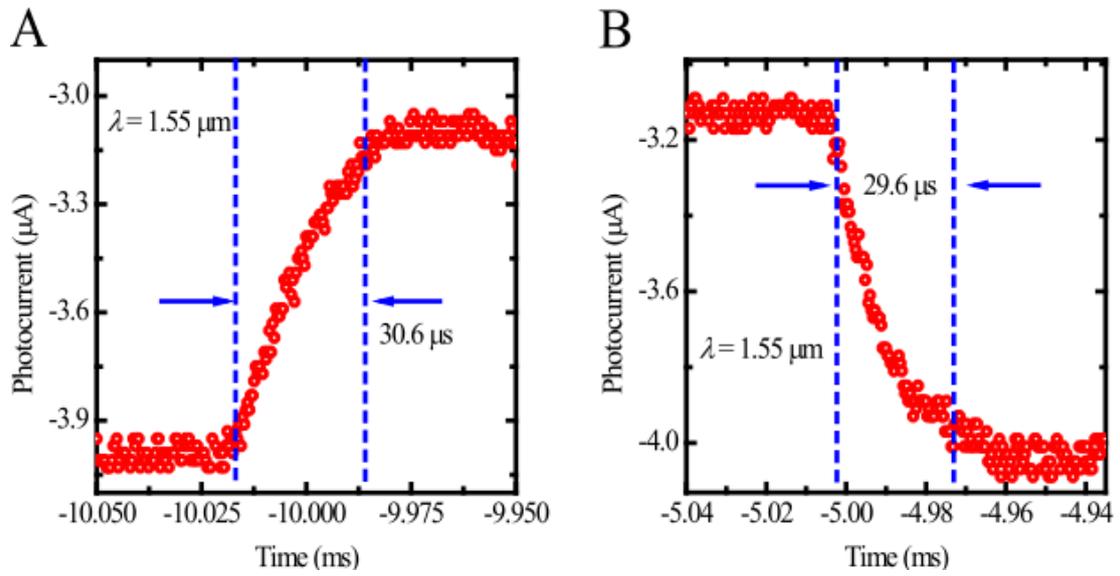

**fig. S5. Fast photoresponse at near infrared.** (**A**) The rise time ~30.6 μs and (**B**) decay time ~29.6 μs of a typical b-As$_{0.83}$P$_{0.17}$ FET measured under a 1550 nm laser at $V_{ds}$ = 0 V and $V_g$ = 0 V in ambient air. The data was collected by an oscilloscope.

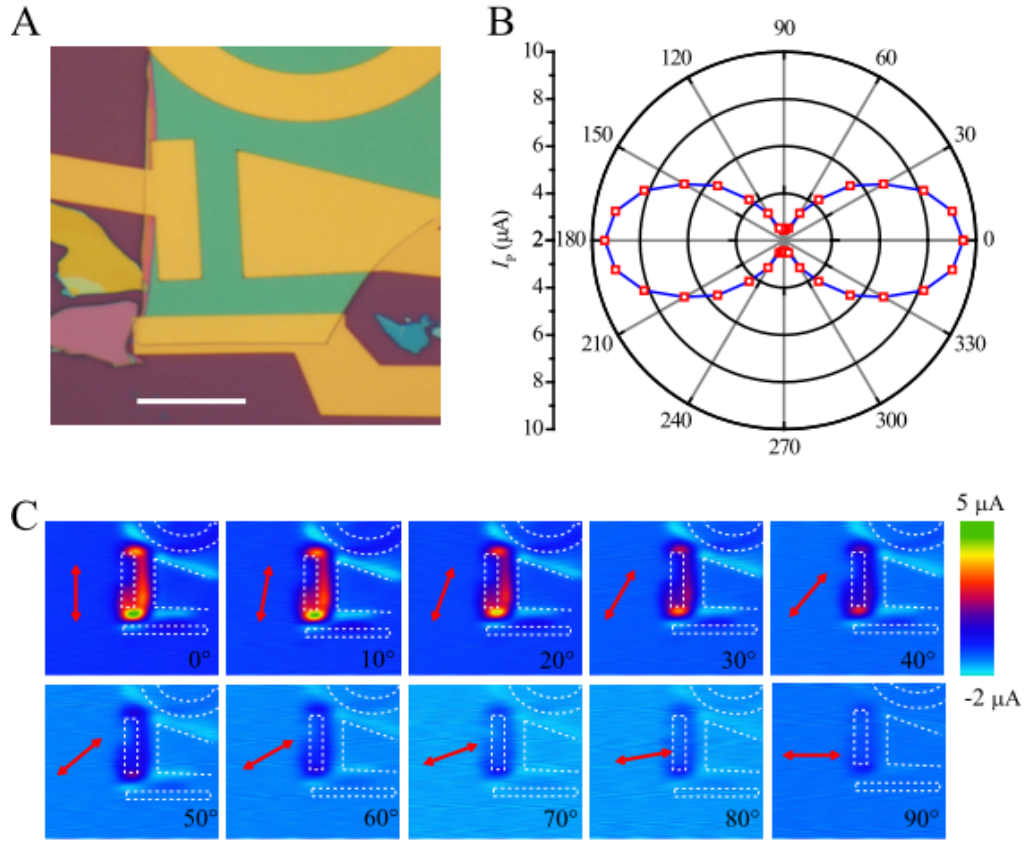

**fig. S6. Laser polarization direction sensitive photocurrent mapping.** (**A**) optical image of a typical b-As$_{0.83}$P$_{0.17}$ FET device. The scale bar is 5 μm. (**B**) The light polarization direction dependent photocurrent at $V_{ds}$=0 V in ambient air. The light polarization direction of the incident laser was adjusted by a half wavelength plate at a step of 10º. The 0º denotes the polarization direction parallel to the contact edge of the metal. The red double arrows present the polarization directions. (**C**) Photocurrent mapping of the b-AsP detector at various laser polarization directions at $V_{ds}$ = 0 V. The wavelength was 1550 nm and the power was fixed at 177.2 μW. The photocurrent is at a maxima with light polarization at 0º and at a minima with light polarization at 90º. The results are in good consistency with the recent studies on black phosphorus.

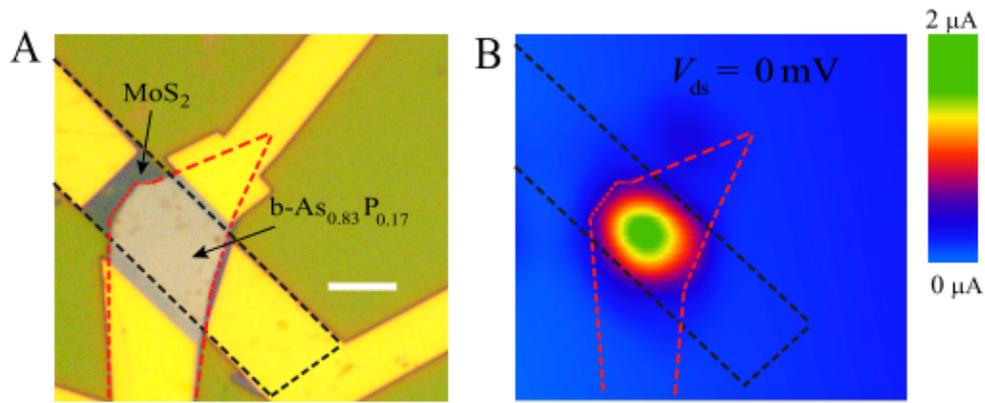

**fig. S7. Photocurrent mapping of the b-As$_{0.83}$P$_{0.17}$/MoS$_2$ heterostructure.** (**A**) The optical image of a typical b-As$_{0.83}$P$_{0.17}$/MoS$_2$ heterostructure device. The scale bar is 5 μm. (**B**) The photocurrent mapping of the b-As$_{0.83}$P$_{0.17}$/MoS$_2$ heterostructure photodetector at $V_{ds}$ = 0 V.

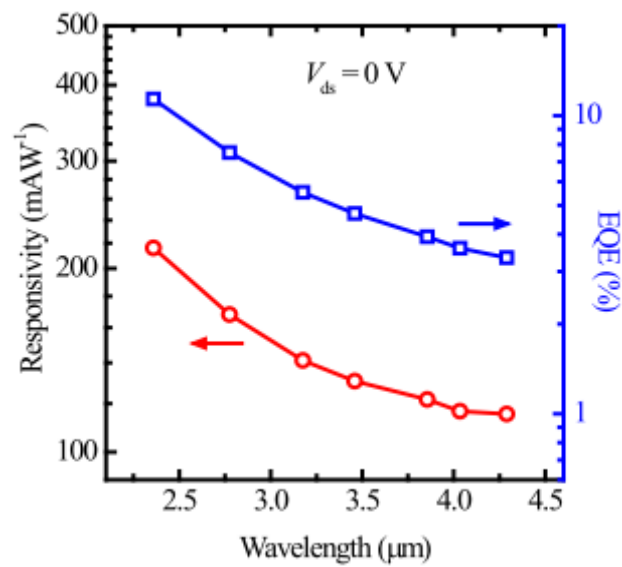

**fig. S8. Photoresponsivity and EQE of a typical b-As$_{0.83}$P$_{0.17}$/MoS$_2$ heterostructure device.** The photoresponsivity decreased from 216.1 mAW$^{-1}$ to 115.4 mAW$^{-1}$ as the wavelength increased from 2.36 μm to 4.29 μm (corresponding EQE decreased from 11.36% to 3.33%). The measurements were carried out at $V_{ds}$ = 0 V and $V_g$ = 0 V.

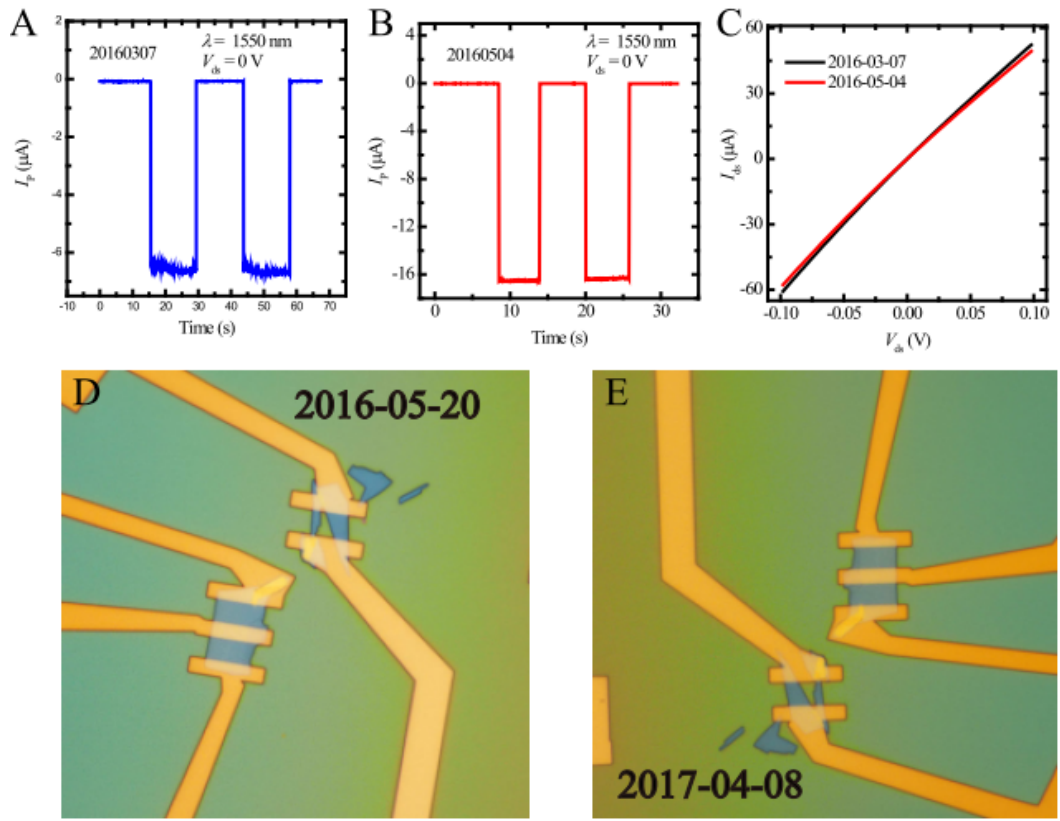

**fig. S9. The stability of b-AsP protected by a PMMA layer.** (A) and (B) The photovoltaic responses of a typical b-AsP FET device at 1.55 μm for the as-fabricated device and after stored in air for two months. (C) The I-V curves of a typical b-AsP FET device. The black and red lines present data from the as-fabricated device and the one after stored in air for two months. (D) and (E) optical images of two as-fabricated devices and the ones after being stored for over ten months.